\documentclass{article}
\usepackage[latin1]{inputenc}

\usepackage[svgnames]{xcolor}
\usepackage{graphicx}
\usepackage{tabularx}
\usepackage{pgfplots}
\usepackage{listings} 
\usepackage{url}
\usepackage{epsfig}
\usepackage{fixltx2e}
\usepackage{sverb}
\usepackage[squaren,Gray]{SIunits}
\usepackage{balance}
\usepackage[colorlinks=true,citecolor=Green]{hyperref}
\usepackage{authblk}
\usepackage{caption}

\usepackage[margin=1in]{geometry}
\pgfplotsset{compat=1.8}

\usepackage[backend=biber,style=authoryear,hyperref=true,backref=true,url=false,doi=true,isbn=false,dashed=false,maxbibnames=99]{biblatex}
\DeclareNameAlias{sortname}{last-first}

\usetikzlibrary{pgfplots.groupplots}
\usepgfplotslibrary{units}

\lstdefinelanguage[ARM]{Assembler}%
  {morekeywords=[1]{.text,.globl,.align},%
  morekeywords=[2]{add, adr, sub, addc, adcs, adds, b, stmdb, %
    sub.W, beq,bge,blt,bne,%
    umull, umlal, cmp, ldr, mov, mul, pop, push, mrs, msr, %
    subs,vadd,vld1,vldm,vmov,vmul,%
    vpadd,vpop,vpush, str, b, bl,%
    itttt, ittee, itt, itte, it???, it??, itee, ittt, addne, subeq, moveq, subeq, movne, eorne, eor, movs,bx,
   },%
   morekeywords=[3]{f32,s32,i32},
   morekeywords=[4]{pc,r0,r1,r2,r3,r4,r5,r6,r7,r8,r9,r10,r11,r12,lr,sp,psp,control,basepri},
   keywordsprefix=.,%
   sensitive=false,%
   morecomment=[l]{;},
   moredelim=*[directive]\#,%
   moredirectives={define,elif,else,endif,error,if,ifdef,ifndef,line,%
      include,pragma,undef,warning}%
  }[keywords,comments,directives]

\definecolor{listing_background}{RGB}{250,250,250}
\definecolor{registers}{rgb}{0,0.4,0}
\definecolor{comments}{rgb}{0.4,0.4,0.4}
\definecolor{CouleurLiens}{cmyk}{1, .50, .10, .01}

\DeclareCaptionFont{white}{ \color{white} }
\DeclareCaptionFormat{listing}{
    \colorbox{CouleurLiens}{
        {\parbox{\textwidth}{#1#2#3}}}
}
\title{Experimental evaluation of two software countermeasures against fault attacks}
\author[1,2]{Nicolas Moro}
\author[2]{Karine Heydemann}
\author[3]{Amine Dehbaoui}
\author[1]{Bruno Robisson}
\author[2]{Emmanuelle Encrenaz}

\affil[1]{CEA, CEA-Tech PACA, LSAS, 13541 Gardanne, France
\authorcr \texttt{nicolas.moro@cea.fr}}
\affil[2]{Sorbonne Universités, UPMC Univ Paris 06, UMR 7606, LIP6, 75005 Paris, France
\authorcr \texttt{karine.heydemann@lip6.fr}}
\affil[3]{SERMA Technologies, CESTI, 33615 Pessac, France
\authorcr \texttt{a.dehbaoui@serma.com}}

\hypersetup{
    pdftitle = {Experimental evaluation of two software countermeasures against fault attacks},
    pdfauthor = {Nicolas Moro, Karine Heydemann, Amine Dehbaoui, Bruno Robisson, Emmanuelle Encrenaz},
    pdfsubject = {2014 IEEE International Symposium on Hardware-Oriented Security and Trust (HOST)},
    pdfkeywords = {microcontroller, fault injection, instruction skip, assembly, countermeasure},
    linkcolor=CouleurLiens, 
    citecolor=CouleurLiens, 
    urlcolor=CouleurLiens 
}

\date{}

\begin{document}

\maketitle

\begin{abstract}
Injection of transient faults can be used as a way to attack embedded systems. On embedded processors such as microcontrollers, several studies showed that such a transient fault injection with glitches or electromagnetic pulses could corrupt either the data loads from the memory or the assembly instructions executed by the circuit. Some countermeasure schemes which rely on temporal redundancy have been proposed to handle this issue. Among them, several schemes add this redundancy at assembly instruction level. In this paper, we perform a practical evaluation for two of those countermeasure schemes by using a pulsed electromagnetic fault injection process on a 32-bit microcontroller. We provide some necessary conditions for an efficient implementation of those countermeasure schemes in practice. We also evaluate their efficiency and highlight their limitations. To the best of our knowledge, no experimental evaluation of the security of such instruction-level countermeasure schemes has been published yet.
\end{abstract}

\makeatletter
\def\blfootnote{\xdef\@thefnmark{}\@footnotetext}
\makeatother

\blfootnote{This work was done while Amine Dehbaoui was with École Nationale Supérieure des Mines de Saint-Étienne (ENSM.SE), 13541 Gardanne, France.}

\blfootnote{This article has been presented at the IEEE International Symposium on Hardware-Oriented Security and Trust (HOST 2014, Arlington, USA). A copy of the presentation can be found on \url{http://www.nicolasmoro.net/research}.}

\blfootnote{© 2014 IEEE. Personal use of this material is permitted. Permission from IEEE must be obtained for all other uses, in any current or future media, including reprinting/republishing this material for advertising or promotional purposes, creating new collective works, for resale or redistribution to servers or lists, or reuse of any copyrighted component of this work in other works. The final publication is available at IEEE via \url{http://dx.doi.org/10.1109/HST.2014.6855580}.}

\section{Introduction} 
Physical attacks were introduced in the late 1990s as a new way to break cryptosystems by exploiting weaknesses in their implementation. Among them, fault attacks were introduced by Boneh \textit{et al.} in 1997 \parencite{Boneh1997}. Those attacks consist in applying a stress to the circuit in order to induce transient faults which could create an attack path \parencite{Barenghi2012}. Such transient faults can be induced in a large set of embedded circuits by using many physical means which include circuit underpowering \parencite{Bhasin2009}, clock glitches \parencite{Balasch2011}, voltage glitches \parencite{Zussa2013}, changes in the temperature \parencite{Skorobogatov2009} or laser shots \parencite{Trichina2010}. More recently, two other fault injection techniques based on using electromagnetic waves have been proposed, either by using a harmonic injection signal \parencite{Poucheret2011} or by using electromagnetic glitches \parencite{Dehbaoui2012}. These physical fault injection means enable to perform higher-level attack schemes such as Differential Fault Analysis (DFA) or safe-error attacks \parencite{Karaklajic2013}. 

Those higher-level attack schemes all rely on an attacker's fault model, which is an abstraction of the set of faults an attacker can perform \parencite{Barenghi2012}. Using such a fault model is necessary to design both software and hardware countermeasures. Defining such an abstracted model requires a good understanding of the effects of the fault injection means. As many kinds of faults can be obtained even with a single fault injection technique, the practical efficiency of a countermeasure highly depends on the accuracy of the considered fault model. Thus, some experiments are necessary both to define realistic fault models and to guarantee the practical efficiency of a countermeasure.

In this paper, we experimentally evaluate the robustness of two software countermeasure schemes against fault injection on embedded programs. These two countermeasures have slightly different purposes and could be combined together. Both of them are designed at assembly code level and rely on providing some replacement sequences to strengthen some sensitive instructions. The first one, proposed in previous works \parencite{Moro2014}, aims at ensuring a fault tolerant execution. It covers almost all the instructions of the considered instruction set and has been formally proven resistant against an instruction skip fault model. The second one was proposed by Barenghi \textit{et al.} \parencite{Barenghi2010}. It uses an instruction duplication approach to perform a fault detection. It has been designed using a more generic fault model but covers a smaller set of instructions. The evaluation experiments that are conducted in this paper will enable us to determine some necessary conditions for an efficient implementation of these countermeasures and to highlight their possible limitations.




The rest of this paper is organized as follows. Section \ref{Section:RelatedWorks} provides an overview of some existing software countermeasure schemes and of the considered injection means for the experiments. Section \ref{Section:ExperimentalSetup} introduces the experimental platform and environment. Section \ref{Section:ExperimentalEvaluation} describes the two studied countermeasures and provides a practical evaluation of their robustness on simple assembly codes. Finally, Section \ref{Section:FreeRTOS} details some results obtained for the two countermeasures on some more complex codes from a FreeRTOS implementation.

\section{Related works}
\label{Section:RelatedWorks}

This section reviews software countermeasures for embedded systems in \ref{Subsection:RelatedWorksCountermeasures} and motivates the use of an electromagnetic fault injection technique for the experiments in \ref{Subsection:RelatedWorksEM}.

\subsection{Software countermeasures}
\label{Subsection:RelatedWorksCountermeasures}

On embedded systems, software-only countermeasure bring some flexibility and avoid any modification on the underlying hardware. Against fault attacks, common countermeasure techniques directly come from software-implemented fault tolerance (SWIFT) techniques \parencite{Reis2005}. Such countermeasure schemes include temporal redundancy, parity checking or checksum-based error detection \parencite{Barenghi2012}. For cryptographic implementations, those SWIFT principles have mostly been applied at a function-level or algorithm-level \parencite{Oboril2013}. Otherwise, some algorithm-specific countermeasures \parencite{Joye2012}, some applicative countermeasures to protect Java Card applets \parencite{Sere2011} or some combined software-hardware countermeasure schemes \parencite{Arora2005} have also been designed.

Those countermeasures are defined with respect to an attacker's model. Such a model provides a theoretical set of faults an attacker could produce. Since performing practical experiments on software countermeasures may require some advanced fault injection means and can be very time-consuming, fault models are also used to perform fault injection simulations \parencite{Theissing2013} or formal proofs \parencite{Moro2014}. Those simulations help to provide stronger guarantees about the efficiency of the tested countermeasures. However, certification processes include practical experiments \parencite{JIL2009}. Thus, the strongest guarantee can only be brought by performing practical experiments on real devices. To the best of our knowledge, no practical evaluation of the efficiency of some generic assembly-level countermeasures has been proposed yet.


\subsection{Electromagnetic fault injection technique}
\label{Subsection:RelatedWorksEM}

Pulsed electromagnetic fault injection has been introduced in the last decade and has turned out to be an effective way to inject transient faults in a circuit's computation. Recent works, such as \parencite{Dehbaoui2012} or \parencite{Moro2013} tend to show that pulsed electromagnetic fault injection could enable to induce faults that are very similar to the faults obtained with clock glitches \parencite{Balasch2011}, voltage glitches \parencite{Zussa2013} or even laser shots on the logic part of a microcontroller \parencite{Trichina2010}. Thus, we think that this electromagnetic fault injection technique should still be representative enough to get a good evaluation of the efficiency of the tested countermeasures.




\section{Experimental setup}
\label{Section:ExperimentalSetup}

\begin{figure}[!ht]
\centering
\includegraphics[scale=0.50]{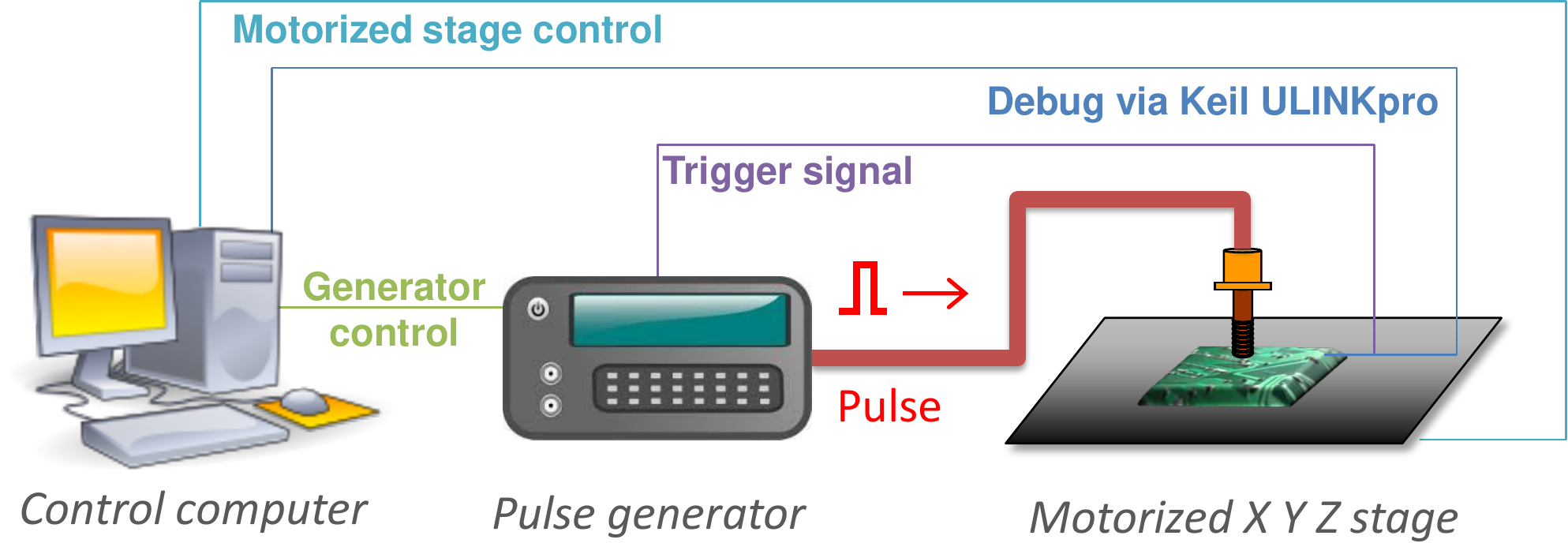}
\caption{Electromagnetic fault injection bench}
\label{Image:Banc}
\end{figure}

\subsection{Pulsed electromagnetic fault injection}
Conductors such as the rails of a power distribution network are one of the primary electromagnetic interferences risk factors for a circuit. They also act as antennas for the radiated electromagnetic pulse flux generated by a coil. This magnetic flux then induces an electromotive force in the power distribution network that leads to a violation of a circuit's timing constraints \parencite{Poucheret2011}. In \parencite{Omarouayache2013}, Omarouayache \textit{et al.} studied magnetic probes built on the basis of small wire loops for the purpose of near-field injection. Their investigation have lead to some useful guidelines to design an electromagnetic antenna. They show that the antenna must be designed as wide-band components to transfer the electromagnetic power with the best efficiency. It was also shown that few loops must be used to optimize the field intensity, and the introduction of a point-sharpened ferrite in the middle of the loop concentrates efficiently the field for near-field operation. Such a magnetic probe enables to induce voltage drops in the target circuit. Those voltage drops then lead to violations of the timing constraints and to faults in the target circuit \parencite{Dehbaoui2012}.



\subsection{Target circuit}
\label{Subsection:Target}
The chosen target is an up-to-date 32-bit microcontroller designed in a CMOS \unit{130}{\nano\meter} technology. It is based on the ARM Cortex-M3 processor \parencite{DefinitiveGuideARMCortexM3}. Its operating frequency is set to \unit{56}{\mega\hertz} without any cache memory. It is also important to mention that no prefetch buffer is activated. Thus, the full execution of some instruction can take several cycles. Cortex-M3 processors use a Harvard architecture and run the ARM Thumb-2 instruction set\footnote{See ARM Architecture Reference Manual - Thumb-2 Supplement, 2005}, which contains both 16-bit and 32-bit instructions. The target circuit embeds some basic security mechanisms against some low-cost fault injection techniques such as clock and voltage glitches. Some interrupt vectors can handle several hardware faults and can be used for a basic fault detection. 

\subsection{Electromagnetic fault injection bench}
Figure \ref{Image:Banc} shows an architectural view of the electromagnetic fault injection platform. It is based on a high speed voltage pulse generator and uses a coil with few turns (diameter of \unit{500}{\micro\meter}) as injection antenna. The pulse generator is used to deliver voltage pulses (from \unit{-210}{\volt} to \unit{210}{\volt}) to the magnetic coil. It has a constant rise and fall transition time of \unit{2}{\nano\second}. For our experiments, the pulses' width is set to \unit{10}{\nano\second}. The target circuit is mounted on a high-accuracy X Y Z motorized stage. The position of the injection antenna is the same for all the experiments of this paper, it has been found by a trial-and-reset approach. This bench also includes some standard control elements such as a PC, an oscilloscope and a Keil ULINKpro debug system. The computer sends pulse injection parameters to the pulse generator. It also controls the target board by using the Keil \micro{}Vision UVSOCK library\footnote{Keil UVSOCK: http://www.keil.com/appnotes/docs/apnt\_{}198.asp}. Since the microcontroller is restarted before injecting a fault, every fault injection attempt requires about \unit{1}{\second}. This fault injection bench and the influence of the different experimental parameters have been presented in more details in previous works \parencite{Moro2013}.

\section{Experimental evaluation of the countermeasures}
\label{Section:ExperimentalEvaluation}

The following section first introduces the faults that can be obtained with our experimental setup in \ref{Subsection:FaultModel} and the approach we use for the evaluation in \ref{Subsection:EvaluationApproach}. Then, it provides an experimental evaluation of the two studied countermeasures on a very simple assembly code in \ref{Subsection:EvaluationFaultTolerance} and \ref{Subsection:EvaluationFaultDetection}.

\subsection{Preliminaries about the fault model}
\label{Subsection:FaultModel}

The fault injection technique we use enables to induce violations of the timing constraints of an integrated circuit \parencite{Dehbaoui2012}. For a microcontroller, bus transfers from the Flash memory are the operations that requires the longest time in a clock cycle \parencite{Moro2013}. Thus, the bus transfers from the Flash memory can easily be corrupted by using delay faults. The Flash memory contains both instructions and data. Thereby, two pipeline stages may be hit by such a technique: the \textit{fetch} stage (for every instruction) and the \textit{decode} stage. In the \textit{fetch} stage, the circuit fetches 32 bits of data from the instruction memory at every clock cycle. Nevertheless, the Thumb-2 instruction set contains both 16-bit and 32-bit instructions. Thus, two instructions at a time might be corrupted. Moreover, \textit{load} instructions also fetch a piece of data in the \textit{decode} phase. Thus, this piece of data can be corrupted too.

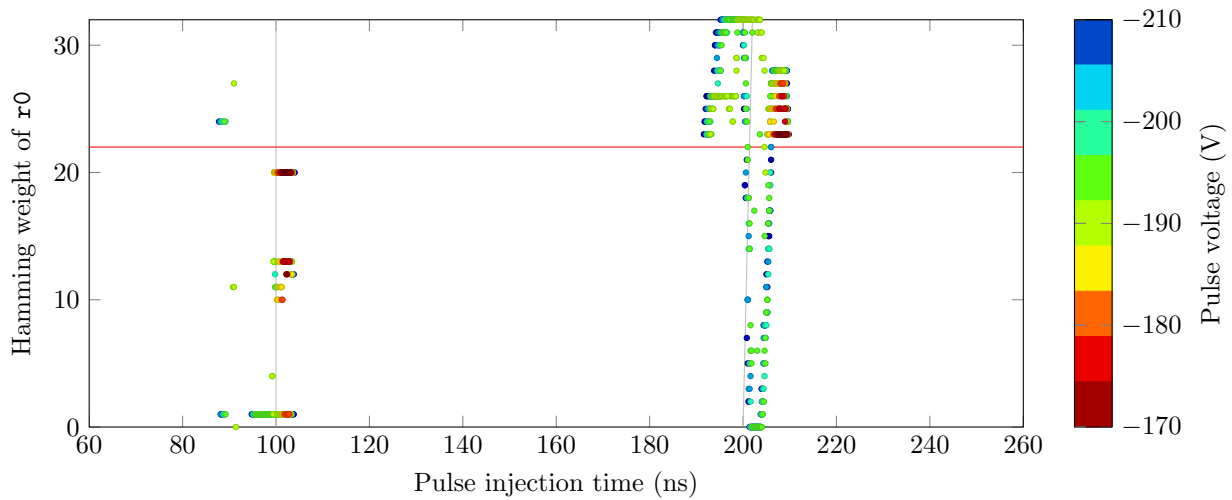
\begin{figure}[!ht]
\centering
\begin{tikzpicture}
\pgfplotsset{height=7cm, width=14cm, compat=newest}
\begin{axis}[scatter, xlabel=Pulse injection time (ns),ylabel=Hamming weight of \texttt{r0},xmin=60,xmax=260,ymin=0,ymax=32,colorbar sampled,colormap/bluered,colorbar style={ylabel={Pulse voltage (V)},samples=10,y dir=reverse},point meta min=-210, point meta max=-170,scatter/@pre marker code/.append style={/tikz/mark size=1}]
\addplot [lightgray, no markers] coordinates {(100,0) (100,32)};
\addplot [lightgray, no markers] coordinates {(200,0) (202,32)};	
	\addplot[only marks,scatter,scatter src=explicit,mark=*]
		coordinates {
(87.8,24) [-210]
(88.2,1) [-210]
(88.2,1) [-210]
(88.8,1) [-210]
(94.8,1) [-210]
(95,1) [-210]
(95.4,1) [-210]
(95.4,1) [-210]
(95.6,1) [-210]
(95.6,1) [-210]
(95.8,1) [-210]
(95.8,1) [-210]
(96,1) [-210]
(96,1) [-210]
(96.2,1) [-210]
(96.2,1) [-210]
(96.4,1) [-210]
(96.4,1) [-210]
(96.6,1) [-210]
(96.6,1) [-210]
(96.8,1) [-210]
(96.8,1) [-210]
(97,1) [-210]
(97,1) [-210]
(97.2,1) [-210]
(97.2,1) [-210]
(97.4,1) [-210]
(97.4,1) [-210]
(97.6,1) [-210]
(97.6,1) [-210]
(97.8,1) [-210]
(97.8,1) [-210]
(98,1) [-210]
(98,1) [-210]
(98.2,1) [-210]
(98.2,1) [-210]
(98.4,1) [-210]
(98.4,1) [-210]
(98.6,1) [-210]
(98.6,1) [-210]
(98.8,1) [-210]
(98.8,1) [-210]
(99,1) [-210]
(99,1) [-210]
(99.2,1) [-210]
(99.2,1) [-210]
(99.4,1) [-210]
(99.4,1) [-210]
(99.6,1) [-210]
(99.6,1) [-210]
(99.8,1) [-210]
(99.8,1) [-210]
(100,1) [-210]
(100,1) [-210]
(100.2,1) [-210]
(100.4,1) [-210]
(103.4,1) [-210]
(103.4,1) [-210]
(103.6,1) [-210]
(103.6,1) [-210]
(103.8,1) [-210]
(103.8,12) [-210]
(104,20) [-210]
(191.6,23) [-210]
(191.8,24) [-210]
(191.8,24) [-210]
(192,25) [-210]
(192.2,26) [-210]
(192.2,26) [-210]
(192.4,23) [-210]
(192.4,26) [-210]
(192.6,24) [-210]
(192.6,26) [-210]
(192.8,26) [-210]
(192.8,25) [-210]
(193,26) [-210]
(193,26) [-210]
(193.2,26) [-210]
(193.2,26) [-210]
(193.4,26) [-210]
(193.4,26) [-210]
(193.6,26) [-210]
(193.6,26) [-210]
(193.8,28) [-210]
(193.8,26) [-210]
(194,30) [-210]
(194,30) [-210]
(194.2,31) [-210]
(194.2,26) [-210]
(194.4,26) [-210]
(194.4,31) [-210]
(194.6,31) [-210]
(194.6,31) [-210]
(194.8,31) [-210]
(194.8,31) [-210]
(195,30) [-210]
(195,31) [-210]
(195.2,32) [-210]
(195.2,31) [-210]
(195.4,32) [-210]
(195.4,31) [-210]
(195.6,32) [-210]
(195.6,32) [-210]
(195.8,31) [-210]
(195.8,31) [-210]
(196,32) [-210]
(196,32) [-210]
(196.2,32) [-210]
(196.2,32) [-210]
(196.4,32) [-210]
(196.4,32) [-210]
(196.6,32) [-210]
(196.6,32) [-210]
(196.8,32) [-210]
(196.8,32) [-210]
(197,32) [-210]
(197,32) [-210]
(197.2,32) [-210]
(197.2,32) [-210]
(197.4,32) [-210]
(197.4,32) [-210]
(197.6,32) [-210]
(197.6,32) [-210]
(197.8,32) [-210]
(197.8,32) [-210]
(198,32) [-210]
(198,32) [-210]
(198.2,32) [-210]
(198.2,32) [-210]
(198.4,32) [-210]
(198.4,32) [-210]
(198.6,32) [-210]
(198.6,32) [-210]
(198.8,32) [-210]
(198.8,32) [-210]
(199,32) [-210]
(199,32) [-210]
(199.2,32) [-210]
(199.2,32) [-210]
(199.4,32) [-210]
(199.4,32) [-210]
(199.6,32) [-210]
(199.6,32) [-210]
(199.8,32) [-210]
(199.8,32) [-210]
(200,32) [-210]
(200,30) [-210]
(200.2,26) [-210]
(200.2,25) [-210]
(200.4,19) [-210]
(200.4,19) [-210]
(200.6,18) [-210]
(200.6,18) [-210]
(200.8,7) [-210]
(200.8,21) [-210]
(201,10) [-210]
(201,5) [-210]
(201.2,2) [-210]
(201.2,2) [-210]
(201.4,0) [-210]
(201.4,5) [-210]
(201.6,0) [-210]
(201.6,0) [-210]
(201.8,0) [-210]
(201.8,0) [-210]
(202,0) [-210]
(202,0) [-210]
(202.2,0) [-210]
(202.2,0) [-210]
(202.4,0) [-210]
(202.4,0) [-210]
(202.6,0) [-210]
(202.6,0) [-210]
(202.8,0) [-210]
(202.8,0) [-210]
(203,0) [-210]
(203,0) [-210]
(203.2,0) [-210]
(203.2,0) [-210]
(203.4,0) [-210]
(203.4,0) [-210]
(203.6,0) [-210]
(203.6,0) [-210]
(203.8,0) [-210]
(203.8,1) [-210]
(204,3) [-210]
(204,3) [-210]
(204.2,2) [-210]
(204.4,8) [-210]
(204.4,5) [-210]
(204.6,8) [-210]
(204.6,8) [-210]
(204.8,8) [-210]
(204.8,8) [-210]
(205,11) [-210]
(205,12) [-210]
(205.2,13) [-210]
(205.2,11) [-210]
(205.4,16) [-210]
(205.4,15) [-210]
(205.6,17) [-210]
(205.6,15) [-210]
(205.8,17) [-210]
(205.8,17) [-210]
(206,20) [-210]
(206,21) [-210]
(206.2,26) [-210]
(206.2,25) [-210]
(206.4,28) [-210]
(206.4,28) [-210]
(206.6,28) [-210]
(206.6,28) [-210]
(206.8,28) [-210]
(206.8,28) [-210]
(207,28) [-210]
(207,28) [-210]
(207.2,28) [-210]
(207.2,28) [-210]
(207.4,28) [-210]
(207.4,28) [-210]
(207.6,28) [-210]
(207.6,28) [-210]
(207.8,28) [-210]
(207.8,28) [-210]
(208,28) [-210]
(208,28) [-210]
(208.2,28) [-210]
(208.2,28) [-210]
(208.4,28) [-210]
(208.4,28) [-210]
(208.6,28) [-210]
(208.6,28) [-210]
(208.8,28) [-210]
(208.8,28) [-210]
(209,28) [-210]
(209,28) [-210]
(209.2,28) [-210]
(209.2,28) [-210]
(209.4,28) [-210]
(209.4,27) [-210]
(209.6,25) [-210]
(209.6,25) [-210]
(209.8,23) [-210]
(88,24) [-205]
(88.2,1) [-205]
(88.4,1) [-205]
(88.4,1) [-205]
(89,1) [-205]
(95,1) [-205]
(95.2,1) [-205]
(95.4,1) [-205]
(95.4,1) [-205]
(95.6,1) [-205]
(95.6,1) [-205]
(95.8,1) [-205]
(95.8,1) [-205]
(96,1) [-205]
(96,1) [-205]
(96.2,1) [-205]
(96.2,1) [-205]
(96.4,1) [-205]
(96.4,1) [-205]
(96.6,1) [-205]
(96.6,1) [-205]
(96.8,1) [-205]
(96.8,1) [-205]
(97,1) [-205]
(97,1) [-205]
(97.2,1) [-205]
(97.2,1) [-205]
(97.4,1) [-205]
(97.4,1) [-205]
(97.6,1) [-205]
(97.6,1) [-205]
(97.8,1) [-205]
(97.8,1) [-205]
(98,1) [-205]
(98,1) [-205]
(98.2,1) [-205]
(98.2,1) [-205]
(98.4,1) [-205]
(98.4,1) [-205]
(98.6,1) [-205]
(98.6,1) [-205]
(98.8,1) [-205]
(98.8,1) [-205]
(99,1) [-205]
(99,1) [-205]
(99.2,1) [-205]
(99.2,1) [-205]
(99.4,1) [-205]
(99.4,1) [-205]
(99.6,1) [-205]
(99.6,1) [-205]
(99.8,1) [-205]
(99.8,1) [-205]
(100,1) [-205]
(100,1) [-205]
(100.2,1) [-205]
(100.4,1) [-205]
(103.2,1) [-205]
(103.2,1) [-205]
(103.4,1) [-205]
(103.4,1) [-205]
(103.6,12) [-205]
(103.6,12) [-205]
(192.2,24) [-205]
(192.2,23) [-205]
(192.4,25) [-205]
(192.8,23) [-205]
(192.8,23) [-205]
(193,26) [-205]
(193,26) [-205]
(193.2,25) [-205]
(193.2,26) [-205]
(193.4,26) [-205]
(193.4,26) [-205]
(193.6,26) [-205]
(193.6,26) [-205]
(193.8,26) [-205]
(193.8,26) [-205]
(194,26) [-205]
(194,26) [-205]
(194.2,26) [-205]
(194.2,26) [-205]
(194.4,29) [-205]
(194.4,28) [-205]
(194.6,30) [-205]
(194.6,30) [-205]
(194.8,31) [-205]
(194.8,31) [-205]
(195,28) [-205]
(195,31) [-205]
(195.2,31) [-205]
(195.2,28) [-205]
(195.4,31) [-205]
(195.4,31) [-205]
(195.6,32) [-205]
(195.6,32) [-205]
(195.8,32) [-205]
(195.8,31) [-205]
(196,31) [-205]
(196,32) [-205]
(196.2,32) [-205]
(196.2,32) [-205]
(196.4,32) [-205]
(196.4,32) [-205]
(196.6,32) [-205]
(196.6,32) [-205]
(196.8,32) [-205]
(196.8,32) [-205]
(197,32) [-205]
(197,32) [-205]
(197.2,32) [-205]
(197.2,32) [-205]
(197.4,32) [-205]
(197.4,32) [-205]
(197.6,32) [-205]
(197.6,32) [-205]
(197.8,32) [-205]
(197.8,32) [-205]
(198,32) [-205]
(198,32) [-205]
(198.2,32) [-205]
(198.2,32) [-205]
(198.4,32) [-205]
(198.4,32) [-205]
(198.6,32) [-205]
(198.6,32) [-205]
(198.8,32) [-205]
(198.8,32) [-205]
(199,32) [-205]
(199,32) [-205]
(199.2,32) [-205]
(199.2,32) [-205]
(199.4,32) [-205]
(199.4,32) [-205]
(199.6,32) [-205]
(199.6,32) [-205]
(199.8,32) [-205]
(199.8,32) [-205]
(200,31) [-205]
(200,31) [-205]
(200.2,26) [-205]
(200.2,29) [-205]
(200.4,24) [-205]
(200.6,29) [-205]
(200.6,20) [-205]
(200.8,18) [-205]
(200.8,18) [-205]
(201,10) [-205]
(201,10) [-205]
(201.2,3) [-205]
(201.2,15) [-205]
(201.4,14) [-205]
(201.4,3) [-205]
(201.6,0) [-205]
(201.6,4) [-205]
(201.8,0) [-205]
(201.8,0) [-205]
(202,0) [-205]
(202,0) [-205]
(202.2,0) [-205]
(202.2,0) [-205]
(202.4,0) [-205]
(202.4,0) [-205]
(202.6,0) [-205]
(202.6,0) [-205]
(202.8,0) [-205]
(202.8,0) [-205]
(203,0) [-205]
(203,0) [-205]
(203.2,0) [-205]
(203.2,0) [-205]
(203.4,0) [-205]
(203.4,0) [-205]
(203.6,0) [-205]
(203.6,0) [-205]
(203.8,0) [-205]
(203.8,1) [-205]
(204,2) [-205]
(204,1) [-205]
(204.2,3) [-205]
(204.2,3) [-205]
(204.4,7) [-205]
(204.4,5) [-205]
(204.6,8) [-205]
(204.6,8) [-205]
(204.8,7) [-205]
(204.8,8) [-205]
(205,9) [-205]
(205,11) [-205]
(205.2,9) [-205]
(205.2,9) [-205]
(205.4,13) [-205]
(205.4,14) [-205]
(205.6,16) [-205]
(205.6,17) [-205]
(205.8,17) [-205]
(205.8,19) [-205]
(206,22) [-205]
(206,22) [-205]
(206.2,28) [-205]
(206.2,27) [-205]
(206.4,27) [-205]
(206.4,28) [-205]
(206.6,28) [-205]
(206.6,28) [-205]
(206.8,28) [-205]
(206.8,28) [-205]
(207,28) [-205]
(207,28) [-205]
(207.2,28) [-205]
(207.2,28) [-205]
(207.4,28) [-205]
(207.4,28) [-205]
(207.6,28) [-205]
(207.6,28) [-205]
(207.8,28) [-205]
(207.8,28) [-205]
(208,28) [-205]
(208,28) [-205]
(208.2,28) [-205]
(208.2,28) [-205]
(208.4,28) [-205]
(208.4,28) [-205]
(208.6,28) [-205]
(208.6,28) [-205]
(208.8,28) [-205]
(208.8,28) [-205]
(209,28) [-205]
(209,28) [-205]
(209.2,28) [-205]
(209.2,28) [-205]
(209.4,25) [-205]
(209.4,26) [-205]
(209.6,24) [-205]
(209.6,24) [-205]
(88.4,24) [-200]
(88.6,24) [-200]
(88.6,24) [-200]
(88.8,1) [-200]
(89.2,24) [-200]
(95.2,1) [-200]
(95.4,1) [-200]
(95.6,1) [-200]
(95.6,1) [-200]
(95.8,1) [-200]
(95.8,1) [-200]
(96,1) [-200]
(96,1) [-200]
(96.2,1) [-200]
(96.2,1) [-200]
(96.4,1) [-200]
(96.4,1) [-200]
(96.6,1) [-200]
(96.6,1) [-200]
(96.8,1) [-200]
(96.8,1) [-200]
(97,1) [-200]
(97,1) [-200]
(97.2,1) [-200]
(97.2,1) [-200]
(97.4,1) [-200]
(97.4,1) [-200]
(97.6,1) [-200]
(97.6,1) [-200]
(97.8,1) [-200]
(97.8,1) [-200]
(98,1) [-200]
(98,1) [-200]
(98.2,1) [-200]
(98.2,1) [-200]
(98.4,1) [-200]
(98.4,1) [-200]
(98.6,1) [-200]
(98.8,1) [-200]
(98.8,1) [-200]
(99,1) [-200]
(99,1) [-200]
(99.2,1) [-200]
(99.2,1) [-200]
(99.4,1) [-200]
(99.4,1) [-200]
(99.6,1) [-200]
(99.6,1) [-200]
(99.8,12) [-200]
(99.8,11) [-200]
(100,11) [-200]
(100,1) [-200]
(100.2,1) [-200]
(100.2,1) [-200]
(100.4,1) [-200]
(103,1) [-200]
(103,1) [-200]
(103.2,1) [-200]
(103.2,1) [-200]
(103.4,13) [-200]
(103.4,12) [-200]
(103.6,20) [-200]
(103.6,20) [-200]
(192.2,23) [-200]
(192.4,24) [-200]
(192.4,24) [-200]
(192.8,25) [-200]
(193,26) [-200]
(193,24) [-200]
(193.2,26) [-200]
(193.2,26) [-200]
(193.4,26) [-200]
(193.4,26) [-200]
(193.6,26) [-200]
(193.6,25) [-200]
(193.8,26) [-200]
(193.8,26) [-200]
(194,26) [-200]
(194,26) [-200]
(194.2,26) [-200]
(194.2,26) [-200]
(194.4,26) [-200]
(194.4,26) [-200]
(194.6,28) [-200]
(194.6,27) [-200]
(194.8,28) [-200]
(194.8,28) [-200]
(195,26) [-200]
(195,30) [-200]
(195.2,26) [-200]
(195.2,26) [-200]
(195.4,31) [-200]
(195.4,31) [-200]
(195.6,31) [-200]
(195.6,31) [-200]
(195.8,31) [-200]
(195.8,31) [-200]
(196,32) [-200]
(196,31) [-200]
(196.2,32) [-200]
(196.2,32) [-200]
(196.4,31) [-200]
(196.4,32) [-200]
(196.6,31) [-200]
(196.6,32) [-200]
(196.8,32) [-200]
(196.8,32) [-200]
(197,32) [-200]
(197.2,32) [-200]
(197.2,32) [-200]
(197.4,32) [-200]
(197.4,32) [-200]
(197.6,32) [-200]
(197.6,32) [-200]
(197.8,32) [-200]
(197.8,32) [-200]
(198,32) [-200]
(198,32) [-200]
(198.2,32) [-200]
(198.2,32) [-200]
(198.4,32) [-200]
(198.4,32) [-200]
(198.6,32) [-200]
(198.6,32) [-200]
(198.8,32) [-200]
(198.8,32) [-200]
(199,32) [-200]
(199,32) [-200]
(199.2,32) [-200]
(199.2,32) [-200]
(199.4,32) [-200]
(199.4,32) [-200]
(199.6,32) [-200]
(199.6,32) [-200]
(199.8,32) [-200]
(199.8,32) [-200]
(200,32) [-200]
(200,32) [-200]
(200.2,32) [-200]
(200.2,30) [-200]
(200.4,32) [-200]
(200.4,32) [-200]
(200.6,32) [-200]
(200.6,29) [-200]
(200.8,26) [-200]
(200.8,25) [-200]
(201,18) [-200]
(201,18) [-200]
(201.2,14) [-200]
(201.2,16) [-200]
(201.4,5) [-200]
(201.4,14) [-200]
(201.6,2) [-200]
(201.6,0) [-200]
(201.8,0) [-200]
(201.8,0) [-200]
(202,0) [-200]
(202,0) [-200]
(202.2,0) [-200]
(202.2,0) [-200]
(202.4,0) [-200]
(202.4,0) [-200]
(202.6,0) [-200]
(202.6,0) [-200]
(202.8,0) [-200]
(202.8,0) [-200]
(203,0) [-200]
(203,0) [-200]
(203.2,0) [-200]
(203.2,0) [-200]
(203.4,0) [-200]
(203.4,0) [-200]
(203.6,0) [-200]
(203.6,0) [-200]
(203.8,0) [-200]
(203.8,0) [-200]
(204,1) [-200]
(204,1) [-200]
(204.2,2) [-200]
(204.2,2) [-200]
(204.4,3) [-200]
(204.4,2) [-200]
(204.6,4) [-200]
(204.6,5) [-200]
(204.8,8) [-200]
(204.8,8) [-200]
(205,9) [-200]
(205,8) [-200]
(205.2,10) [-200]
(205.2,9) [-200]
(205.4,12) [-200]
(205.4,14) [-200]
(205.6,14) [-200]
(205.6,14) [-200]
(205.8,19) [-200]
(205.8,20) [-200]
(206,25) [-200]
(206,27) [-200]
(206.2,27) [-200]
(206.2,27) [-200]
(206.4,27) [-200]
(206.4,27) [-200]
(206.6,28) [-200]
(206.6,28) [-200]
(206.8,28) [-200]
(206.8,28) [-200]
(207,28) [-200]
(207,28) [-200]
(207.2,28) [-200]
(207.2,28) [-200]
(207.4,28) [-200]
(207.4,28) [-200]
(207.6,28) [-200]
(207.6,28) [-200]
(207.8,28) [-200]
(207.8,28) [-200]
(208,28) [-200]
(208,28) [-200]
(208.2,28) [-200]
(208.2,28) [-200]
(208.4,28) [-200]
(208.4,28) [-200]
(208.6,28) [-200]
(208.6,28) [-200]
(208.8,28) [-200]
(208.8,28) [-200]
(209,28) [-200]
(209,28) [-200]
(209.2,26) [-200]
(209.2,28) [-200]
(209.4,25) [-200]
(209.4,25) [-200]
(209.6,23) [-200]
(209.6,23) [-200]
(89,24) [-195]
(89.2,1) [-195]
(95.6,1) [-195]
(95.6,1) [-195]
(95.8,1) [-195]
(96,1) [-195]
(96,1) [-195]
(96.2,1) [-195]
(96.2,1) [-195]
(96.4,1) [-195]
(96.4,1) [-195]
(96.6,1) [-195]
(96.6,1) [-195]
(96.8,1) [-195]
(97,1) [-195]
(97,1) [-195]
(97.2,1) [-195]
(97.2,1) [-195]
(97.4,1) [-195]
(97.4,1) [-195]
(97.6,1) [-195]
(97.6,1) [-195]
(97.8,1) [-195]
(97.8,1) [-195]
(98,1) [-195]
(98,1) [-195]
(98.2,1) [-195]
(98.4,1) [-195]
(98.4,1) [-195]
(98.6,1) [-195]
(98.8,1) [-195]
(98.8,1) [-195]
(99,1) [-195]
(99,1) [-195]
(99.2,1) [-195]
(99.2,1) [-195]
(99.4,1) [-195]
(99.4,1) [-195]
(99.6,13) [-195]
(99.6,13) [-195]
(99.8,13) [-195]
(99.8,13) [-195]
(100,11) [-195]
(100,11) [-195]
(100.2,1) [-195]
(100.2,1) [-195]
(100.4,1) [-195]
(100.4,1) [-195]
(100.6,1) [-195]
(100.6,1) [-195]
(100.8,1) [-195]
(100.8,1) [-195]
(101,1) [-195]
(101,1) [-195]
(101.2,1) [-195]
(101.2,1) [-195]
(101.4,1) [-195]
(101.4,1) [-195]
(101.6,1) [-195]
(101.6,1) [-195]
(102,1) [-195]
(102.2,1) [-195]
(102.8,1) [-195]
(103,1) [-195]
(103,1) [-195]
(103.2,1) [-195]
(103.2,1) [-195]
(103.4,12) [-195]
(103.4,13) [-195]
(103.6,20) [-195]
(192.6,23) [-195]
(192.6,23) [-195]
(192.8,24) [-195]
(193,25) [-195]
(193,25) [-195]
(193.2,23) [-195]
(193.2,26) [-195]
(193.4,26) [-195]
(193.4,26) [-195]
(193.6,26) [-195]
(193.6,26) [-195]
(193.8,26) [-195]
(194,26) [-195]
(194,26) [-195]
(194.2,26) [-195]
(194.2,26) [-195]
(194.4,26) [-195]
(194.4,26) [-195]
(194.6,26) [-195]
(194.6,26) [-195]
(194.8,26) [-195]
(194.8,26) [-195]
(195,26) [-195]
(195,26) [-195]
(195.2,26) [-195]
(195.2,28) [-195]
(195.4,30) [-195]
(195.4,26) [-195]
(195.6,31) [-195]
(195.6,31) [-195]
(195.8,31) [-195]
(195.8,31) [-195]
(196,31) [-195]
(196,31) [-195]
(196.2,26) [-195]
(196.2,31) [-195]
(196.4,31) [-195]
(196.4,26) [-195]
(196.6,32) [-195]
(196.6,32) [-195]
(196.8,32) [-195]
(196.8,32) [-195]
(197,32) [-195]
(197,32) [-195]
(197.2,32) [-195]
(197.2,32) [-195]
(197.4,26) [-195]
(197.4,26) [-195]
(197.6,32) [-195]
(197.6,26) [-195]
(197.8,32) [-195]
(197.8,32) [-195]
(198,32) [-195]
(198,32) [-195]
(198.2,32) [-195]
(198.2,26) [-195]
(198.4,32) [-195]
(198.4,26) [-195]
(198.6,29) [-195]
(198.6,32) [-195]
(198.8,31) [-195]
(198.8,32) [-195]
(199,32) [-195]
(199,32) [-195]
(199.2,32) [-195]
(199.2,32) [-195]
(199.4,32) [-195]
(199.4,32) [-195]
(199.6,32) [-195]
(199.6,32) [-195]
(199.8,32) [-195]
(199.8,32) [-195]
(200,32) [-195]
(200,32) [-195]
(200.2,32) [-195]
(200.2,32) [-195]
(200.4,29) [-195]
(200.4,31) [-195]
(200.6,25) [-195]
(200.6,27) [-195]
(200.8,24) [-195]
(200.8,24) [-195]
(201,22) [-195]
(201,21) [-195]
(201.2,18) [-195]
(201.2,18) [-195]
(201.4,14) [-195]
(201.4,16) [-195]
(201.6,6) [-195]
(201.6,8) [-195]
(201.8,6) [-195]
(201.8,5) [-195]
(202,0) [-195]
(202,31) [-195]
(202.2,32) [-195]
(202.2,0) [-195]
(202.4,17) [-195]
(202.4,0) [-195]
(202.6,0) [-195]
(202.6,0) [-195]
(202.8,0) [-195]
(202.8,0) [-195]
(203,6) [-195]
(203,0) [-195]
(203.2,0) [-195]
(203.2,31) [-195]
(203.4,0) [-195]
(203.4,0) [-195]
(203.6,0) [-195]
(203.6,23) [-195]
(203.8,0) [-195]
(203.8,0) [-195]
(204,0) [-195]
(204,0) [-195]
(204.2,1) [-195]
(204.2,1) [-195]
(204.4,2) [-195]
(204.4,2) [-195]
(204.6,6) [-195]
(204.6,15) [-195]
(204.8,5) [-195]
(204.8,7) [-195]
(205,9) [-195]
(205,9) [-195]
(205.2,10) [-195]
(205.2,10) [-195]
(205.4,19) [-195]
(205.4,16) [-195]
(205.6,17) [-195]
(205.6,18) [-195]
(205.8,26) [-195]
(205.8,24) [-195]
(206,27) [-195]
(206,27) [-195]
(206.2,27) [-195]
(206.2,27) [-195]
(206.4,27) [-195]
(206.4,27) [-195]
(206.6,27) [-195]
(206.6,27) [-195]
(206.8,28) [-195]
(206.8,28) [-195]
(207,28) [-195]
(207,28) [-195]
(207.2,28) [-195]
(207.2,27) [-195]
(207.4,28) [-195]
(207.4,28) [-195]
(207.6,28) [-195]
(207.6,28) [-195]
(207.8,28) [-195]
(207.8,28) [-195]
(208,28) [-195]
(208,28) [-195]
(208.2,28) [-195]
(208.2,28) [-195]
(208.4,28) [-195]
(208.4,28) [-195]
(208.6,28) [-195]
(208.6,28) [-195]
(208.8,28) [-195]
(208.8,28) [-195]
(209,28) [-195]
(209,27) [-195]
(209.2,25) [-195]
(209.2,26) [-195]
(209.4,25) [-195]
(209.4,25) [-195]
(209.6,24) [-195]
(209.6,23) [-195]
(90.8,11) [-190]
(91,27) [-190]
(91,11) [-190]
(91.4,0) [-190]
(99.2,4) [-190]
(99.2,4) [-190]
(99.4,13) [-190]
(99.4,1) [-190]
(99.6,20) [-190]
(99.6,20) [-190]
(99.8,20) [-190]
(99.8,20) [-190]
(100,11) [-190]
(100,11) [-190]
(100.2,11) [-190]
(100.2,1) [-190]
(100.4,1) [-190]
(100.4,11) [-190]
(100.6,1) [-190]
(100.6,13) [-190]
(100.8,11) [-190]
(100.8,1) [-190]
(101,1) [-190]
(101,1) [-190]
(101.2,1) [-190]
(101.2,1) [-190]
(101.4,1) [-190]
(101.4,1) [-190]
(101.6,1) [-190]
(101.6,1) [-190]
(101.8,1) [-190]
(101.8,1) [-190]
(102,1) [-190]
(102,1) [-190]
(102.2,1) [-190]
(102.2,1) [-190]
(102.4,1) [-190]
(102.4,1) [-190]
(102.6,1) [-190]
(102.6,1) [-190]
(102.8,1) [-190]
(102.8,1) [-190]
(103,1) [-190]
(103,1) [-190]
(103.2,1) [-190]
(103.2,13) [-190]
(103.4,13) [-190]
(103.4,12) [-190]
(193,23) [-190]
(193.2,25) [-190]
(193.6,25) [-190]
(193.6,25) [-190]
(193.8,26) [-190]
(193.8,26) [-190]
(194.2,26) [-190]
(194.2,26) [-190]
(194.4,26) [-190]
(194.8,26) [-190]
(194.8,26) [-190]
(195,26) [-190]
(195,26) [-190]
(195.2,26) [-190]
(195.2,26) [-190]
(195.4,26) [-190]
(195.4,26) [-190]
(195.6,26) [-190]
(196,26) [-190]
(196,26) [-190]
(196.2,26) [-190]
(196.2,26) [-190]
(196.4,26) [-190]
(196.6,26) [-190]
(196.6,26) [-190]
(197,25) [-190]
(197,25) [-190]
(197.2,25) [-190]
(197.8,24) [-190]
(197.8,24) [-190]
(198,26) [-190]
(198,26) [-190]
(198.2,26) [-190]
(198.2,26) [-190]
(198.4,26) [-190]
(198.6,28) [-190]
(198.6,29) [-190]
(198.8,32) [-190]
(199,32) [-190]
(199,32) [-190]
(199.2,32) [-190]
(199.2,32) [-190]
(199.4,32) [-190]
(199.4,32) [-190]
(199.6,32) [-190]
(199.8,32) [-190]
(199.8,32) [-190]
(200,32) [-190]
(200.2,32) [-190]
(200.2,32) [-190]
(200.4,32) [-190]
(200.4,32) [-190]
(200.6,32) [-190]
(200.8,32) [-190]
(201,32) [-190]
(201,32) [-190]
(201.2,32) [-190]
(201.4,32) [-190]
(201.4,32) [-190]
(201.6,32) [-190]
(201.6,32) [-190]
(201.8,32) [-190]
(201.8,32) [-190]
(202,32) [-190]
(202,32) [-190]
(202.2,32) [-190]
(202.4,32) [-190]
(202.8,32) [-190]
(203,32) [-190]
(203.2,31) [-190]
(203.2,32) [-190]
(203.4,32) [-190]
(203.4,32) [-190]
(203.6,32) [-190]
(203.6,32) [-190]
(203.8,31) [-190]
(204,31) [-190]
(204,29) [-190]
(204.4,29) [-190]
(204.6,28) [-190]
(204.6,22) [-190]
(204.8,20) [-190]
(205.2,23) [-190]
(205.2,23) [-190]
(205.4,23) [-190]
(205.4,25) [-190]
(205.6,25) [-190]
(205.6,25) [-190]
(205.8,25) [-190]
(205.8,25) [-190]
(206,25) [-190]
(206,25) [-190]
(206.2,27) [-190]
(206.2,27) [-190]
(206.4,27) [-190]
(206.4,27) [-190]
(206.6,25) [-190]
(206.6,26) [-190]
(206.8,27) [-190]
(206.8,27) [-190]
(207,27) [-190]
(207,27) [-190]
(207.2,27) [-190]
(207.2,27) [-190]
(207.4,27) [-190]
(207.4,27) [-190]
(207.6,27) [-190]
(207.6,27) [-190]
(207.8,28) [-190]
(207.8,28) [-190]
(208,28) [-190]
(208,28) [-190]
(208.2,27) [-190]
(208.2,28) [-190]
(208.4,28) [-190]
(208.4,28) [-190]
(208.6,28) [-190]
(208.6,27) [-190]
(208.8,28) [-190]
(208.8,27) [-190]
(209,27) [-190]
(209,27) [-190]
(209.2,25) [-190]
(209.2,25) [-190]
(209.4,24) [-190]
(209.4,25) [-190]
(209.6,23) [-190]
(209.6,23) [-190]
(99.6,20) [-185]
(99.6,20) [-185]
(99.8,20) [-185]
(100,20) [-185]
(100,20) [-185]
(100.2,20) [-185]
(100.2,10) [-185]
(100.4,10) [-185]
(100.4,10) [-185]
(100.6,20) [-185]
(100.6,13) [-185]
(100.8,20) [-185]
(100.8,13) [-185]
(101,11) [-185]
(101,11) [-185]
(101.2,11) [-185]
(101.2,11) [-185]
(101.4,13) [-185]
(101.4,1) [-185]
(101.6,1) [-185]
(101.6,1) [-185]
(101.8,1) [-185]
(101.8,1) [-185]
(102,1) [-185]
(102,1) [-185]
(102.2,1) [-185]
(102.2,1) [-185]
(102.4,1) [-185]
(102.4,1) [-185]
(102.6,1) [-185]
(102.6,1) [-185]
(102.8,1) [-185]
(102.8,1) [-185]
(103,1) [-185]
(103,1) [-185]
(103.2,12) [-185]
(103.2,20) [-185]
(103.4,20) [-185]
(103.4,12) [-185]
(103.6,20) [-185]
(103.6,20) [-185]
(205.4,23) [-185]
(205.4,23) [-185]
(205.6,23) [-185]
(205.6,23) [-185]
(205.8,24) [-185]
(205.8,24) [-185]
(206,25) [-185]
(206,25) [-185]
(206.2,25) [-185]
(206.2,25) [-185]
(206.4,23) [-185]
(206.4,25) [-185]
(206.6,25) [-185]
(206.6,24) [-185]
(206.8,25) [-185]
(206.8,25) [-185]
(207,25) [-185]
(207,26) [-185]
(207.2,25) [-185]
(207.2,27) [-185]
(207.4,27) [-185]
(207.4,25) [-185]
(207.6,27) [-185]
(207.6,27) [-185]
(207.8,27) [-185]
(207.8,27) [-185]
(208,27) [-185]
(208,27) [-185]
(208.2,27) [-185]
(208.2,27) [-185]
(208.4,27) [-185]
(208.4,27) [-185]
(208.6,27) [-185]
(208.6,27) [-185]
(208.8,27) [-185]
(208.8,27) [-185]
(209,27) [-185]
(209,25) [-185]
(209.2,25) [-185]
(209.2,25) [-185]
(209.4,23) [-185]
(209.4,23) [-185]
(100.4,20) [-180]
(100.4,20) [-180]
(100.6,20) [-180]
(100.6,20) [-180]
(100.8,20) [-180]
(100.8,20) [-180]
(101,20) [-180]
(101,20) [-180]
(101.2,20) [-180]
(101.2,10) [-180]
(101.4,13) [-180]
(101.4,10) [-180]
(101.6,13) [-180]
(101.6,13) [-180]
(101.8,13) [-180]
(101.8,1) [-180]
(102,13) [-180]
(102,13) [-180]
(102.2,1) [-180]
(102.2,1) [-180]
(102.4,1) [-180]
(102.4,1) [-180]
(102.6,1) [-180]
(102.6,1) [-180]
(102.8,1) [-180]
(102.8,1) [-180]
(103,13) [-180]
(103,13) [-180]
(103.2,20) [-180]
(103.2,20) [-180]
(103.4,20) [-180]
(103.4,20) [-180]
(206.6,23) [-180]
(206.6,23) [-180]
(206.8,23) [-180]
(207,25) [-180]
(207,23) [-180]
(207.2,23) [-180]
(207.2,23) [-180]
(207.4,25) [-180]
(207.4,25) [-180]
(207.6,25) [-180]
(207.6,25) [-180]
(207.8,26) [-180]
(207.8,26) [-180]
(208,27) [-180]
(208,25) [-180]
(208.2,26) [-180]
(208.2,27) [-180]
(208.4,26) [-180]
(208.4,27) [-180]
(208.6,27) [-180]
(208.6,27) [-180]
(208.8,26) [-180]
(208.8,26) [-180]
(209,25) [-180]
(209,25) [-180]
(209.2,25) [-180]
(209.2,25) [-180]
(209.4,23) [-180]
(209.4,23) [-180]
(209.6,23) [-180]
(101,20) [-175]
(101,20) [-175]
(101.2,20) [-175]
(101.2,20) [-175]
(101.4,20) [-175]
(101.4,20) [-175]
(101.6,13) [-175]
(101.6,13) [-175]
(101.8,20) [-175]
(101.8,13) [-175]
(102,13) [-175]
(102,13) [-175]
(102.2,13) [-175]
(102.2,12) [-175]
(102.4,13) [-175]
(102.4,13) [-175]
(102.6,13) [-175]
(102.6,13) [-175]
(102.8,13) [-175]
(102.8,13) [-175]
(103,20) [-175]
(103,20) [-175]
(103.2,20) [-175]
(207,23) [-175]
(207.4,23) [-175]
(207.6,23) [-175]
(207.6,23) [-175]
(207.8,25) [-175]
(207.8,25) [-175]
(208,25) [-175]
(208,25) [-175]
(208.2,23) [-175]
(208.2,25) [-175]
(208.4,26) [-175]
(208.4,26) [-175]
(208.6,25) [-175]
(208.6,23) [-175]
(208.8,25) [-175]
(208.8,25) [-175]
(209,24) [-175]
(209,25) [-175]
(209.2,23) [-175]
(209.2,23) [-175]
(209.4,23) [-175]
(101.4,20) [-170]
(101.4,20) [-170]
(101.6,20) [-170]
(101.6,20) [-170]
(101.8,20) [-170]
(101.8,20) [-170]
(102,20) [-170]
(102,20) [-170]
(102.2,20) [-170]
(102.2,20) [-170]
(102.4,12) [-170]
(102.4,20) [-170]
(102.6,20) [-170]
(102.6,20) [-170]
(102.8,20) [-170]
(102.8,20) [-170]
(103,20) [-170]
(103,20) [-170]
(207.4,23) [-170]
(207.6,23) [-170]
(207.8,23) [-170]
(208,23) [-170]
(208,23) [-170]
(208.2,23) [-170]
(208.2,23) [-170]
(208.4,23) [-170]
(208.4,23) [-170]
(208.6,23) [-170]
(208.6,23) [-170]
(208.8,23) [-170]
(208.8,23) [-170]
(209,23) [-170]
(209,23) [-170]
(209.2,23) [-170]
(209.2,23) [-170]};

\addplot [red, no markers] coordinates {(60,22) (260,22)};

\end{axis}
\end{tikzpicture}
\caption{Fault injection results for a \texttt{ldr r0,=0xCAFECAFE} instruction}
\label{Figure:Resultats-Load}
\end{figure} 

To highlight these two trends, we performed a fault injection experiment on a single 16-bit \texttt{ldr} instruction. In this example, the target instruction is a \texttt{ldr r0,[pc,\#40]} PC-relative load instruction that has been generated by the \texttt{armasm} assembler from the macro \texttt{ldr r0,=0xCAFECAFE}. This loaded value has been chosen because it is a very specific value that cannot been found anywhere else in the memory or cannot be the output of almost all the possible instruction replacements. Moreover, some of the hardware exception handlers enable to access the address of the instruction that triggered the exception. We use this piece of information to define the time intervals that will be swept in our experiments. For this experiment, we performed a fault injection for 9 voltage values (from \unit{-210}{\volt} to \unit{-170}{\volt}) and over a \unit{200}{\nano\second} time interval. Another analysis we performed with positive voltages led to the same patterns for this target instruction. Such a \unit{200}{\nano\second} time interval is very long in comparison to the \unit{17.8}{\nano\second} clock period. This can be explained both by the fact that we need to cover the three pipeline stages of the instruction and by the fact that several clock cycles are necessary for the memory fetches. This time interval has been swept by steps of \unit{200}{\pico\second}. For every injection time, the fault injection process has been performed two times. The fault injection results are presented on Fig. \ref{Figure:Resultats-Load}. This graph shows the Hamming weight of the output values in \texttt{r0} when no exception has been triggered depending on the pulse voltage and the pulse injection time. The Hamming weight of the \texttt{0xCAFECAFE} expected value is 22. On this figure, we can clearly distinguish two groups of output faults, located around two injection times: those around around \unit{100}{\nano\second} and those around \unit{200}{\nano\second}. Their distribution of faulty values is very different. The first one corresponds to the \textit{fetch} stage. Since the instruction \textit{opcode} is corrupted, very few instruction corruptions lead to a valid new instruction. Most of them generate an invalid instruction that triggers an exception. Among the instruction \textit{fetch} corruptions that led to a fault in \texttt{r0}, some instructions have also been transformed into branch instructions. The second group corresponds to the \textit{decode} phase. Since the loaded word is corrupted, we can observe a much bigger diversity in the faulty outputs for this injection time. One has to note that such kind of results are not specific to electromagnetic glitches and have been obtained for other fault injection means \parencite{Balasch2011,Trichina2010}.

\subsection{Evaluation approach}
\label{Subsection:EvaluationApproach}

We need to define a relevant metric to evaluate the efficiency of the countermeasures. Since the countermeasure sequences add some instructions, the time to execute a full countermeasure sequence becomes longer than the time to execute the initial instruction. Thus, the number of vulnerable points, \textit{i.e.} the injection times for which a fault injection attempt is successful, may also increase. Comparing the percentage of faulty outputs could appear to be a solution to compare two data sets with different numbers of measurements. Nevertheless, we assume that the most meaningful metric for such a comparison is the number of faults that have been obtained on the destination register. The countermeasure is really effective if it can overcome the fact that some new vulnerable points are added and if it can decrease this number of vulnerable points. Thus, comparing the number of faulty outputs is probably the most relevant metric for an attacker since it indicates the number of potential vulnerabilities on an embedded code. 

Moreover, it is important to mention that we analyzed several metrics such as the global number of faults on any register or the number of faults that match an instruction skip for every experiment. It happened very frequently that several metrics show exactly the same pattern. For clarity purposes on the curves presented in this paper, we chose the metrics we assume to be the most significant for each experiment when the other relevant metrics showed the same trend.

\subsection{Fault tolerance countermeasure}
\label{Subsection:EvaluationFaultTolerance}

This countermeasure aims at providing a fault-tolerant replacement sequence for most of the instruction of an instruction set \parencite{Moro2014}. Such a countermeasure has been designed to be tolerant to any single instruction skip and does not provide any protection to the data flow. An example of the use of this countermeasure (from \cite{Moro2014}) is given in Listing \ref{Listing:FaultTolerance}. In this example, the replacement sequence mimics the effect of a \texttt{bl} instruction by putting the return pointer into the link register \texttt{lr} and branching to the destination function. Even if no fault is injected, the subroutine is only called once, since the return pointer is set after the two \texttt{b} instructions.

\begin{lstlisting}[language={[ARM]Assembler},caption={Fault tolerance countermeasure for a \texttt{bl function} instruction},label={Listing:FaultTolerance}]
    adr  r1, return_label
    adr  r1, return_label
    add  lr, r1, #1  ; Thumb mode requires the
    add  lr, r1, #1  ; last bit of LR to be set
    b    function
    b    function
return_label
\end{lstlisting}

In Listing \ref{Listing:FaultTolerance}, the two \texttt{adr} and the two \texttt{b} instructions are encoded with their 16-bit encoding by default. In order to force the \texttt{armasm} assembler to use a 32-bit encoding, one can use the \texttt{.w} suffix after the instruction mnemonic (\texttt{adr.w}) or use the registers \texttt{r8} to \texttt{r14}. Indeed, since the two \texttt{add} instructions use the \texttt{lr} (\texttt{r14}) register, they are necessarily encoded with a 32-bit size.

Thus, we performed some fault injection experiments on four codes: a \texttt{bl} instruction without countermeasure (\unit{100}{\nano}{\second} time interval by steps of \unit{200}{\pico\second}), a \texttt{bl.w} instruction with forced 32-bit encoding without countermeasure (\unit{100}{\nano}{\second}), the replacement sequence from Listing \ref{Listing:FaultTolerance} (\unit{400}{\nano}{\second}) and this replacement sequence with forced 32-bit encoding (\unit{400}{\nano}{\second}). Several values were used for the pulse voltage, from \unit{-210}{\volt} to \unit{-170}{\volt} and from \unit{120}{\volt} to \unit{150}{\volt} by steps of \unit{5}{\volt}. The target circuit crashes for voltages over \unit{150}{\volt}, and no faults were obtained for pulse voltages between \unit{-160}{\volt} and \unit{120}{\volt}. In this experiment, the subroutine that is called only modifies \texttt{r0}. Thus, we analyze the number of faults in \texttt{r0} at the end of the experiment to evaluate the countermeasure. The \textit{faults on any register} curves correspond to an output in which at least one register contains a value different from the expected one.

\begin{figure}[!ht]
\begin{tikzpicture}
  \begin{axis}[
    name=a,
    legend style={legend pos=north west,font=\tiny},
    legend columns=2,
    legend cell align=left,
    width=.55\linewidth,
    height=6cm,
    axis x line=box,
    axis y line*=left,
    xmin=-210, xmax=-170,
    ymin=0, ymax=250,
	xlabel=Pulse amplitude (V),
	ylabel=Number of faulty outputs,
	grid=major
    ]
	\addplot[smooth,mark=square*,red] plot coordinates {
	(-210,59)
	(-205,61)
	(-200,60)
	(-195,59)
	(-190,56)
	(-185,56)
	(-180,29)
	(-175,0)
	(-170,0)};
	\addlegendentry{No CM, 16-bit, faults on \texttt{r0}}
	

	\addplot[smooth,dotted,mark=square*,red, mark size=1pt] plot coordinates {
	(-210,143)
	(-205,138)
	(-200,150)
	(-195,148)
	(-190,142)
	(-185,143)
	(-180,73)
	(-175,0)
	(-170,0)};
	\addlegendentry{No CM, 16-bit, faults on any register}

	\addplot[smooth,mark=diamond*,orange] plot coordinates {
	(-210,0)
	(-205,0)
	(-200,0)
	(-195,0)
	(-190,0)
	(-185,0)
	(-180,0)
	(-175,0)
	(-170,0)};	
	\addlegendentry{No CM, 32-bit, faults on \texttt{r0}}
	

	\addplot[smooth,dotted,mark=diamond*,orange,mark size=1pt] plot coordinates {	
	(-210,0)
	(-205,0)
	(-200,0)
	(-195,0)
	(-190,0)
	(-185,0)
	(-180,0)
	(-175,0)
	(-170,0)};	
	\addlegendentry{No CM, 32-bit, faults on any register}

	\addplot[smooth,mark=triangle*,blue] plot coordinates {
	(-210,104)
	(-205,92)
	(-200,91)
	(-195,46)
	(-190,20)
	(-185,31)
	(-180,9)
	(-175,15)
	(-170,20)};
	\addlegendentry{FT CM, 16-bit, faults on \texttt{r0}}
	

	\addplot[smooth,dotted,mark=triangle*,blue, mark size=1pt] plot coordinates {
	(-210,123)
	(-205,112)
	(-200,119)
	(-195,73)
	(-190,34)
	(-185,50)
	(-180,24)
	(-175,28)
	(-170,52)};
	\addlegendentry{FT CM, 16-bit, faults on any register}	
	
	\addplot[smooth,mark=*,green] plot coordinates {
	(-210,0)
	(-205,1)
	(-200,0)
	(-195,0)
	(-190,0)
	(-185,0)
	(-180,0)
	(-175,1)
	(-170,2)};
	\addlegendentry{FT CM, 32-bit, faults on \texttt{r0}}


	\addplot[smooth,dotted,mark=*,green, mark size=1pt] plot coordinates {
	(-210,64)
	(-205,69)
	(-200,66)
	(-195,64)
	(-190,39)
	(-185,6)
	(-180,2)
	(-175,14)
	(-170,17)};
	\addlegendentry{FT CM, 32-bit, faults on any register}

  \end{axis}

  \begin{axis}[
    name=b,
    width=.45\linewidth,
    height=6cm,
    xshift=9cm,
    axis x line=box,
    axis y line*=right,    
    xmin=120, xmax=150,
    ymin=0, ymax=250,
	xticklabels={,120,130,140,150},
	yticklabels={},	
	grid=major    
    ]

	\addplot[smooth,mark=square*,red] plot coordinates {
	(120,0)
	(125,0)
	(130,6)
	(135,37)
	(140,43)
	(145,43)
	(150,59)	};


	\addplot[smooth,dotted,mark=square*,red,mark size=1pt] plot coordinates {
	(120,0)
	(125,0)
	(130,9)	
	(135,38)		
	(140,44)		
	(145,45)		
	(150,60)		
	};	

	\addplot[smooth,mark=diamond*,orange] plot coordinates {
	(120,0)
	(125,0)
	(130,0)
	(135,0)
	(140,33)
	(145,38)	
	(150,43)};


	\addplot[smooth,dotted,mark=diamond*,orange,mark size=1pt] plot coordinates {
	(120,0)
	(125,0)
	(130,0)
	(135,0)
	(140,34)
	(145,38)	
	(150,43)};

	\addplot[smooth,mark=triangle*,blue] plot coordinates {
	(120,0)
	(125,0)
	(130,77)
	(135,96)
	(140,112)
	(145,123)	
	(150,132)};
	

	\addplot[smooth,dotted,mark=triangle*,blue,mark size=1pt] plot coordinates {
	(120,0)
	(125,0)
	(130,86)
	(135,104)
	(140,125)
	(145,133)	
	(150,151)};	
	
	\addplot[smooth,mark=*,green] plot coordinates {
	(120,0)
	(125,0)
	(130,3)
	(135,1)	
	(140,1)
	(145,1)
	(150,2)};

	\addplot[smooth,dotted,mark=*,green,mark size=1pt] plot coordinates {
	(120,0)
	(125,0)
	(130,0)
	(135,0)
	(140,21)
	(145,12)	
	(150,3)};	

  \end{axis}

  \draw [dotted] (a.south east) -- (b.south west);
  \draw [dotted] (a.north east) -- (b.north west);
\end{tikzpicture}
\caption{Fault injection results for the fault tolerance countermeasure}
\label{Figure:Resultats-FaultTolerance}
\end{figure}
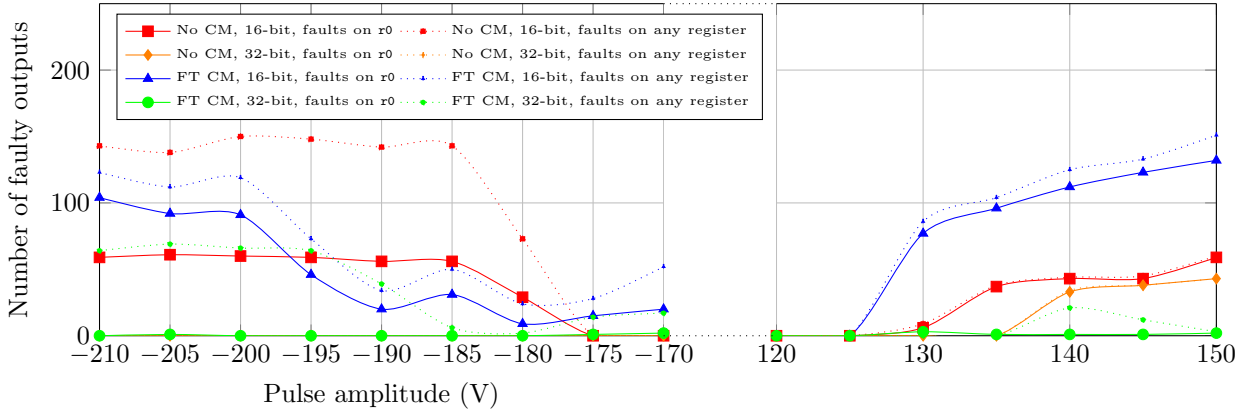

The fault injection results are shown on Fig. \ref{Figure:Resultats-FaultTolerance}. First, we can observe that the curves related to the number of faults in \texttt{r0} and the number of faults on at least one register follow the same patterns, which tends to prove the relevance of choosing the number of faulty outputs in \texttt{r0} as a relevant metric. Then, we can also observe that applying the countermeasure without forced 32-bit encoding does not seem efficient. Indeed, this countermeasure has not been designed to be resistant to two consecutive instruction corruptions. Because of the memory alignment in the experiment, in the replacement sequence the first two \texttt{adr} instructions and later the last two \texttt{b} instructions are loaded in a single \textit{fetch} stage. Thus, it seems that such double corruption happened. Moreover, we can also observe that no faulty output has been obtained for pulses with a negative voltage on a single 32-bit \texttt{bl.w} instruction. Nevertheless, such an instruction could still be faulted by using pulses with a positive voltage. An explanation for such result can be found in the way instructions are encoded. The 16-bit subset of the instruction set is very compact (most of the 16-bit values correspond to one instruction) while the 32-bit subset is very sparse: very few bit flips can change a 16-bit instruction into another instruction, but this assertion is not true for a 32-bit encoding. Finally, for the experiment with a fault tolerance countermeasure and a forced 32-bit encoding, some faults on other registers that are due to instruction \textit{fetch} corruptions and very few faults on \texttt{r0} have still be obtained. To sum up, this countermeasure appears to be very effective for both positive and negative glitches. Applying the countermeasure scheme with a forced 32-bit encoding is a necessary condition to guarantee its efficiency.

\subsection{Fault detection countermeasure}
\label{Subsection:EvaluationFaultDetection}

This countermeasure aims at detecting any single fault, including instruction skips, some cases of instruction replacements and faults on the data flow. It is based on duplicating the execution of an instruction and storing its results in another extra register \parencite{Barenghi2010}. Then, a comparison is done to detect any difference between the two destination registers. If an error is detected, the program branches to an error handler subroutine. This countermeasure can directly be applied to several ALU instructions. Nevertheless, it is not yet applicable to some more special instructions such as branch instructions or instructions that use the flags. As an example, the countermeasure for a \texttt{ldr} instruction (from \cite{Barenghi2010}) is given in Listing \ref{Listing:FaultDetection}. In this code example, a value is loaded from the Flash memory. The address of this value is relative to the program counter.

\begin{lstlisting}[language={[ARM]Assembler},caption={Fault detection countermeasure for a \texttt{ldr} instruction},label={Listing:FaultDetection}]
ldr    r0, [pc, #40] ; initial load instruction
ldr    r1, [pc, #38] ; duplicated load instruction
cmp    r0, r1        ; comparison between r0 and r1
bne    error         ; if r0 != r1, raise an error
\end{lstlisting}

We performed some fault injection experiments on four codes: a single \texttt{ldr} instruction that loads \texttt{0xCAFECAFE} (\unit{150}{\nano\second} time interval by steps of \unit{200}{\pico\second}), a single \texttt{ldr.w} instruction with a forced 32-bit encoding (\unit{150}{\nano\second}), the replacement sequence presented in Listing \ref{Listing:FaultDetection} (\unit{300}{\nano\second}) and the same replacement sequence with a forced 32-bit encoding for every instruction (\unit{500}{\nano\second}).

\begin{figure}[!ht]
\begin{tikzpicture}
  \begin{axis}[
    name=a,
    legend style={legend pos=north west,font=\tiny},
    legend columns=2,
    legend cell align=left,
    width=.55\linewidth,
    height=7cm,
    axis x line=box,
    axis y line*=left,
    xmin=-210, xmax=-170,
    ymin=0, ymax=800,
	xlabel=Pulse amplitude (V),
	ylabel=Number of faulty outputs,
	grid=major
    ]

	\addplot[smooth, mark=square*,red] plot coordinates {
	(-210,243)
	(-205,235)
	(-200,234)
	(-195,240)
	(-190,176)
	(-185,83)
	(-180,62)
	(-175,44)
	(-170,35)};
	\addlegendentry{No CM, 16-bit, faults on \texttt{r0}}
	
	\addplot[smooth,mark=diamond*,orange] plot coordinates {
	(-210,247)
	(-205,245)
	(-200,237)
	(-195,234)
	(-190,232)
	(-185,217)
	(-180,154)
	(-175,52)
	(-170,46)};
	\addlegendentry{No CM, 32-bit, faults on \texttt{r0}}
	
	\addplot[smooth,mark=triangle*,blue] plot coordinates {
	(-210,253)
	(-205,245)
	(-200,253)
	(-195,212)
	(-190,207)
	(-185,202)
	(-180,146)
	(-175,78)
	(-170,56)};
	\addlegendentry{FD CM, 16-bit, faults on \texttt{r0}}

	\addplot[smooth,dashed,mark=triangle*,blue,mark size=1pt] plot coordinates {
	(-210,426)
	(-205,420)
	(-200,424)
	(-195,386)
	(-190,369)
	(-185,360)
	(-180,271)
	(-175,138)
	(-170,99)};
	\addlegendentry{FD CM, 16-bit, detected faults}

	\addplot[smooth,mark=*,green] plot coordinates {
	(-210,2)
	(-205,3)
	(-200,5)
	(-195,0)
	(-190,0)
	(-185,0)
	(-180,0)
	(-175,0)
	(-170,0)};
	\addlegendentry{FD CM, 32-bit, faults on \texttt{r0}}
	
	\addplot[smooth,dashed,mark=*,green, mark size=1pt] plot coordinates {
	(-210,518)
	(-205,508)
	(-200,498)
	(-195,393)
	(-190,357)
	(-185,312)
	(-180,271)
	(-175,141)
	(-170,100)};
	\addlegendentry{FD CM, 32-bit, detected faults}

  \end{axis}

  \begin{axis}[
    name=b,
    width=.45\linewidth,
    height=7cm,
    xshift=9cm,
    axis x line=box,
    axis y line*=right,    
    xmin=120, xmax=150,
    ymin=0, ymax=800,
	xticklabels={,120,130,140,150},
	yticklabels={},
    grid=major
    ]

	\addplot[smooth,mark=square*,red] plot coordinates {
	(120,0)
	(125,0)
	(130,98)
	(135,192)
	(140,205)
	(145,210)	
	(150,233)};

	\addplot[smooth,mark=diamond*,orange] plot coordinates {
	(120,0)
	(125,0)
	(130,118)
	(135,176)
	(140,184)
	(145,200)	
	(150,223)};

	\addplot[smooth,mark=triangle*,blue] plot coordinates {
	(120,0)
	(125,0)
	(130,0)
	(135,2)
	(140,5)
	(145,2)
	(150,1)
	};
	
	\addplot[smooth,dashed,mark=triangle*,blue,mark size=1pt] plot coordinates {
	(120,0)
	(125,0)
	(130,50)
	(135,311)
	(140,332)
	(145,344)
	(150,384)
	};	
	
	\addplot[smooth,mark=*,green] plot coordinates {
	(120,0)
	(125,0)
	(130,2)
	(135,1)
	(140,4)
	(145,1)	
	(150,3)
	};

	\addplot[smooth,dashed,mark=*,green,mark size=1pt] plot coordinates {
	(120,0)
	(125,0)
	(130,267)
	(135,369)
	(140,397)
	(145,420)	
	(150,485)	
	};

  \end{axis}

  \draw [dotted] (a.south east) -- (b.south west);
  \draw [dotted] (a.north east) -- (b.north west);
\end{tikzpicture}

\caption{Fault injection results for the fault detection countermeasure}
\label{Figure:Resultats-FaultDetection}
\end{figure}
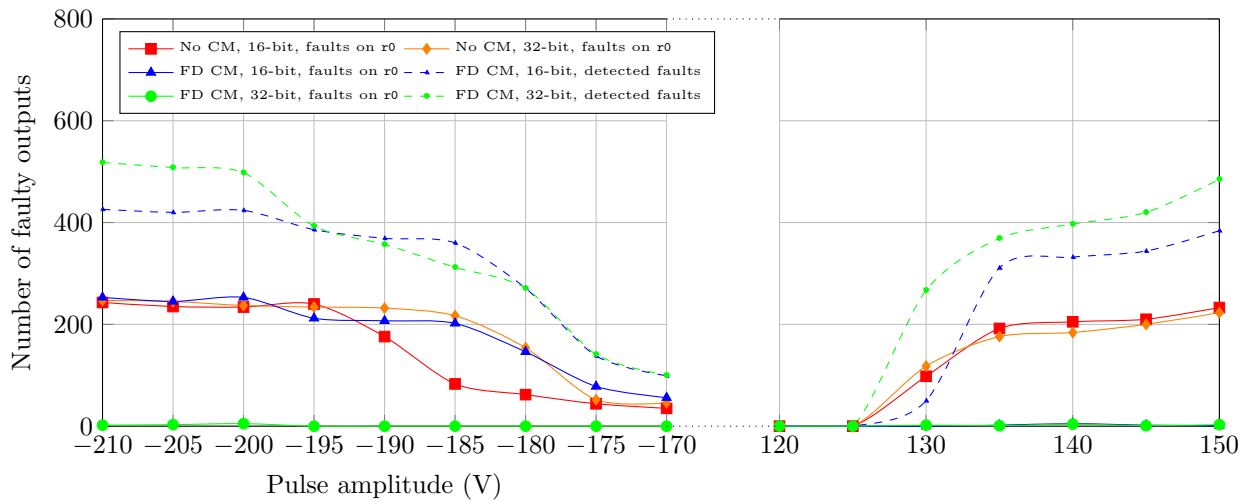 

The fault injection results are presented in Fig. \ref{Figure:Resultats-FaultDetection}. The \textit{detected faults} curves show the number of calls to the \texttt{error} subroutine. From a black box approach, we can observe that applying the countermeasure without forced 32-bit encoding creates more vulnerable injection times than the initial single \texttt{ldr} instruction and does not bring any security for negative glitches. Nevertheless, this countermeasure seems to work very well for positive glitches. Finally, the countermeasure scheme appears to be very effective with a 32-bit encoding of the instructions, it can handle both types of faults presented in \ref{Subsection:FaultModel}, either on the data flow or the control flow.

\section{Evaluation of the countermeasures on a FreeRTOS implementation}
\label{Section:FreeRTOS}

\subsection{FreeRTOS and target implementation}
\label{Subsection:FreeRTOS-Intro}

This section gives details about some experimental results that have been obtained on a FreeRTOS-MPU implementation. FreeRTOS is a portable open-source real-time operating system (RTOS) for embedded devices. It it written in C and has been designed to be very simple to provide a convenient set of tools to design real-time applications. In the following experiments, the FreeRTOS-MPU implementation we use has been built with the Keil MDK-ARM compiler. FreeRTOS is a multitasking operating system. It uses a scheduler to decide which task should be executing. At every interrupt from the system timer, this scheduler gives processing time to the task with the highest priority. FreeRTOS-MPU is a special port of FreeRTOS that uses the Memory Protection Unit (MPU) and the hardware privilege levels of the Cortex-M3 processor. It is able to create tasks in either privileged or unprivileged mode.

We chose to use a FreeRTOS implementation to show the practical interest such countermeasures could have for real-world complex projects that cannot be directly designed in assembly language. Indeed, these countermeasures could directly be applied to a compiled binary to reinforce an embedded system's resistance to fault attacks. A use case for the fault tolerance countermeasure is shown in \ref{Subsection:FreeRTOS-Tolerance}, another one for the fault detection countermeasure is provided in \ref{Subsection:FreeRTOS-Detection}.

\subsection{Fault tolerance countermeasure}
\label{Subsection:FreeRTOS-Tolerance}

At system initialization, the tasks are created and the processor runs in privileged mode. Then, before starting the first task, the \texttt{prvRestoreContextOfFirstTask} function is called. This function uses several types of instructions and sets the execution context to run the first task. It also switches the processor to unprivileged mode if the first task is an unprivileged task. If the systems runs no privileged task, the processor never switches back to privileged mode since the scheduler also runs in unprivileged mode. In particular, this function is theoretically vulnerable to a fault attack: an instruction skip attack can skip the \texttt{msr}\footnote{\texttt{msr} moves the contents of a general-purpose register into a special register} instruction that switches to unprivileged mode. The most sensitive part of this function is shown on Listing \ref{Listing:FreeRTOS-FaultTolerance}.

\begin{lstlisting}[language={[ARM]Assembler},caption={End of the \texttt{prvRestoreContextOfFirstTask} function},label={Listing:FreeRTOS-FaultTolerance}]
msr control, r3    ; switches to unprivileged mode
msr psp, r0        ; initializes the stack pointer
mov r0, #0
msr basepri, r0    ; base priority mask register
ldr lr, =0xfffffffd 
bx  lr             ; returns to Thread mode
\end{lstlisting}

\begin{figure}[!ht]
\begin{tikzpicture}
  \begin{axis}[
    name=a,
    legend style={legend pos=north west,font=\tiny},
    legend columns=2,
	legend cell align=left,    
    width=.55\linewidth,
    height=7cm,
    axis x line=box,
    axis y line*=left,
    xmin=-210, xmax=-170,
    ymin=0, ymax=600,
	xlabel=Pulse amplitude (V),
	ylabel=Number of faulty outputs,
	grid=major
    ]

	\addplot[smooth, mark=square*,red] plot coordinates {
	(-210,3)
	(-205,4)
	(-200,7)
	(-195,4)	
	(-190,7)
	(-185,0)	
	(-180,0)
	(-175,1)
	(-170,3)};
	\addlegendentry{No CM, 16-bit, faults on \texttt{CONTROL}}

	\addplot[smooth,dotted,mark=square*,red,mark size=1pt] plot coordinates {
	(-210,387)
	(-205,383)
	(-200,392)
	(-195,385)	
	(-190,236)
	(-185,209)	
	(-180,214)
	(-175,217)
	(-170,207)};
	\addlegendentry{No CM, 16-bit, faults on any reg.}

	\addplot[smooth,mark=*,green] plot coordinates {
	(-210,28)
	(-205,16)
	(-200,5)
	(-195,28)	
	(-190,26)
	(-185,7)	
	(-180,15)
	(-175,27)
	(-170,25)};
	\addlegendentry{FT CM, 32-bit, faults on \texttt{CONTROL}}

	\addplot[smooth,dotted,mark=*,green,mark size=1pt] plot coordinates {
	(-210,382)
	(-205,357)
	(-200,337)
	(-195,380)	
	(-190,230)
	(-185,185)	
	(-180,144)
	(-175,158)
	(-170,140)};
	\addlegendentry{FT CM, 32-bit, faults on any reg.}
	
  \end{axis}

  \begin{axis}[
    name=b,
    width=.45\linewidth,
    height=7cm,
    xshift=9cm,
    axis x line=box,
    axis y line*=right,    
    xmin=120, xmax=150,
    ymin=0, ymax=600,
	xticklabels={,120,130,140,150},
	yticklabels={},
    grid=major
    ]

	\addplot[smooth,mark=square*,red] plot coordinates {
	(120,0)
	(130,53)
	(135,48)
	(140,66)
	(145,83)
	(150,89)};

	\addplot[smooth,dotted,mark=square*,red,mark size=1pt] plot coordinates {
	(120,0)
	(130,205)	
	(135,268)		
	(140,293)		
	(145,315)		
	(150,320)		
	};	

	\addplot[smooth,mark=*,green] plot coordinates {
	(120,0)
	(130,7)
	(135,24)
	(140,14)
	(145,27)
	(150,37)};

	\addplot[smooth,dotted,mark=*,green,mark size=1pt] plot coordinates {
	(120,0)
	(130,80)	
	(135,150)		
	(140,135)		
	(145,130)		
	(150,182)		
	};	

  \end{axis}

  \draw [dotted] (a.south east) -- (b.south west);
  \draw [dotted] (a.north east) -- (b.north west);
\end{tikzpicture}
\caption{Fault injection results on the \texttt{prvRestoreContextOfFirstTask} function of a FreeRTOS implementation}
\label{Figure:FreeRTOS-FaultTolerance}
\end{figure}
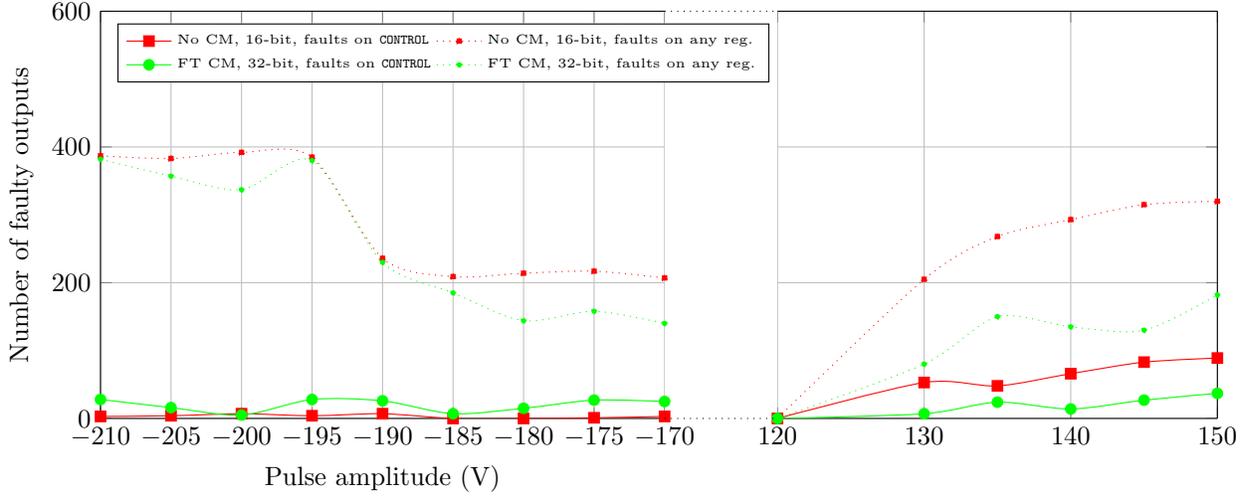

We performed some fault injection experiments on the whole function, either without countermeasure (\unit{2}{\micro\second} time interval) or with the fault tolerance countermeasure and a forced 32-bit encoding applied to every instruction (\unit{5}{\micro\second}). The results are presented on Fig. \ref{Figure:FreeRTOS-FaultTolerance}. They show a very mixed efficiency for the countermeasure on this code, with a good efficiency only for positive glitches. Such a result might be explained by the fact that this countermeasure has been designed for an instruction skip fault model. For some parts of the tested code, this fault model may be too simplified and may be an incorrect abstraction for the injected faults. Indeed, the instruction skip fault model has been observed for different experimental configurations on several targets \parencite{Dehbaoui2012,Barenghi2012}. Nevertheless, some recent research papers have shown that instruction skips could be a specific case of replacements in the instruction binary code \parencite{Balasch2011,Moro2013}. The instruction skip fault model can probably be a good abstraction on simple codes but visibly lacks of relevance for more complex codes. Thus, defining a more accurate fault model seems to be a prerequisite for any future improvement of this countermeasure.

\subsection{Fault detection countermeasure}
\label{Subsection:FreeRTOS-Detection}

Every task has its own level of priority. A constant named \texttt{configMAX\_{}PRIORITIES} is used to set the maximal level of priority that tasks can take. If the system tries to create a task with a higher level of priority, the task is created with the priority level contained in \texttt{configMAX\_{}PRIORITIES}. Moreover, the \texttt{xTaskCreateRestricted} function takes as input a structure which contains the task parameters and uses the content of this structure to call the \texttt{xTaskGenericCreate} function. In such a structure, the \texttt{uxPriority} integer is used to define the task's priority. Since for the tasks that are created at the system's initialization those configuration structures are generally stored in the Flash memory, the \texttt{uxPriority} integer is generally loaded with a \texttt{ldr} instruction. Thus, a fault injection attempt may corrupt this \texttt{ldr} instruction and result into a priority elevation for a specific task. The target code is presented in Listing \ref{Listing:FreeRTOS-FaultDetection}.

\begin{lstlisting}[language={[ARM]Assembler},caption={Last instructions before the call to \texttt{xTaskGenericCreate}},label={Listing:FreeRTOS-FaultDetection}]
ldr  r0, [r0, #0]    ; loads uxPriority in r0
str  r0, [sp, #0]    ; puts uxPriority on the stack
movs r3, #0          ; null pointer (parameters)
movs r2, #128        ; stack depth for the task
movs r1, #0          ; empty string (task's name)
ldr  r0, =address_task_function
bl   xTaskGenericCreate
\end{lstlisting}

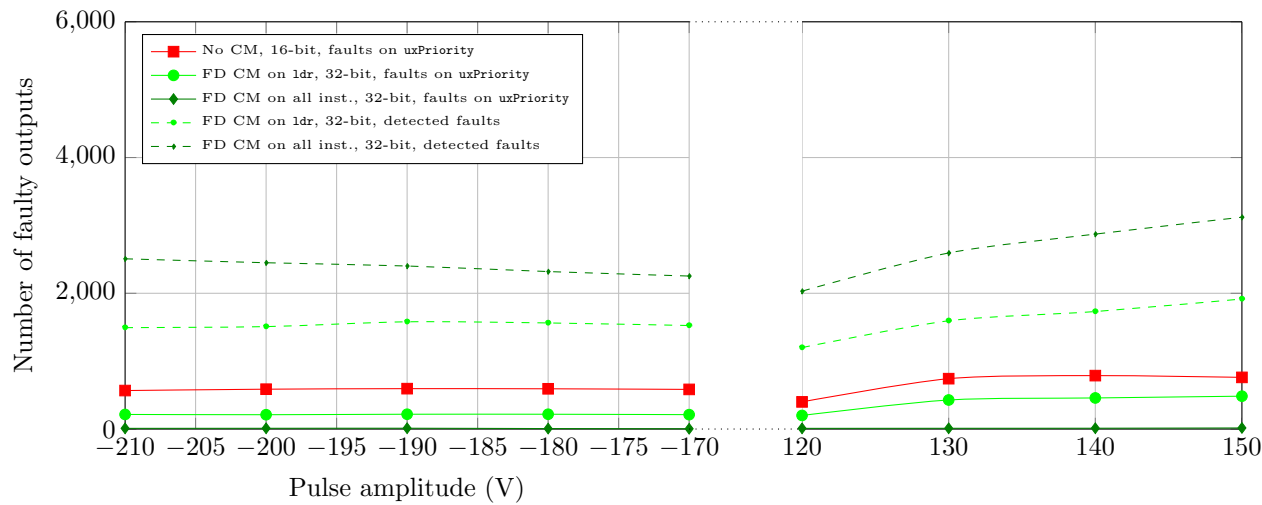
\begin{figure}[!ht]
\begin{tikzpicture}
  \begin{axis}[
    name=a,
    legend style={legend pos=north west,font=\tiny},
    legend columns=1,
    legend cell align=left,
    width=.55\linewidth,
    height=7cm,
    axis x line=box,
    axis y line*=left,
    xmin=-210, xmax=-170,
    ymin=0, ymax=6000,
	xlabel=Pulse amplitude (V),
	ylabel=Number of faulty outputs,
	grid=major
    ]

	\addplot[smooth, mark=square*,red] plot coordinates {
	(-210,567)
	(-200,587)
	(-190,595)
	(-180,593)
	(-170,584)};
	\addlegendentry{No CM, 16-bit, faults on \texttt{uxPriority}}

	\addplot[smooth,mark=*,green] plot coordinates {
	(-210,215)
	(-200,211)
	(-190,218)
	(-180,218)
	(-170,212)};
	\addlegendentry{FD CM on \texttt{ldr}, 32-bit, faults on \texttt{uxPriority}}

	\addplot[smooth,mark=diamond*,Green] plot coordinates {
	(-210,10)
	(-200,11)
	(-190,12)
	(-180,7)
	(-170,4)};
	\addlegendentry{FD CM on all inst., 32-bit, faults on \texttt{uxPriority}}

	\addplot[smooth,dashed,mark=*,green, mark size=1pt] plot coordinates {
	(-210,1494)
	(-200,1509)
	(-190,1580)
	(-180,1563)
	(-170,1525)};
	\addlegendentry{FD CM on \texttt{ldr}, 32-bit, detected faults}

	\addplot[smooth,dashed,mark=diamond*,Green, mark size=1pt] plot coordinates {
	(-210,2508)
	(-200,2450)
	(-190,2402)
	(-180,2320)
	(-170,2254)};
	\addlegendentry{FD CM on all inst., 32-bit, detected faults}

  \end{axis}

  \begin{axis}[
    name=b,
    width=.45\linewidth,
    height=7cm,
    xshift=9cm,
    axis x line=box,
    axis y line*=right,    
    xmin=120, xmax=150,
    ymin=0, ymax=6000,
	xticklabels={,120,130,140,150},
	yticklabels={},
    grid=major
    ]

	\addplot[smooth,mark=square*,red] plot coordinates {
	(120,400)
	(130,742)
	(140,787)
	(150,761)};

	\addplot[smooth,mark=*,green] plot coordinates {
	(120,200)
	(130,428)
	(140,458)
	(150,484)};

	\addplot[smooth,mark=diamond*,Green] plot coordinates {
	(120,10)
	(130,11)
	(140,12)
	(150,16)};

	\addplot[smooth,dashed,mark=*,green, mark size=1pt] plot coordinates {
	(120,1200)
	(130,1595)
	(140,1731)
	(150,1916)};

	\addplot[smooth,dashed,mark=diamond*,Green, mark size=1pt] plot coordinates {
	(120,2031)
	(130,2594)
	(140,2872)
	(150,3121)};

  \end{axis}

  \draw [dotted] (a.south east) -- (b.south west);
  \draw [dotted] (a.north east) -- (b.north west);
\end{tikzpicture}
\caption{Fault injection results on the instructions before the call to the \texttt{xTaskGenericCreate} function of a FreeRTOS implementation}
\label{Figure:FreeRTOS-FaultDetection}
\end{figure}

We performed some fault injection experiments on the 14 assembly instructions (which include 3 \texttt{ldr} instructions) that set the input arguments for the function, either without countermeasure (\unit{1}{\micro\second} time interval) or with the fault detection countermeasure and a forced 32-bit encoding applied only on the \texttt{ldr} instructions (\unit{2}{\micro\second}) or with the fault detection countermeasure and a forced 32-bit encoding on all the instructions (\unit{4}{\micro\second}). The results are presented on Fig. \ref{Figure:FreeRTOS-FaultDetection}. They show an average efficiency when only applied to \texttt{ldr} instructions. The remaining faulty outputs are due to the corruption of other unprotected instructions. Indeed, the countermeasure is very effective when applied to every instruction: less than 20 faulty outputs have been obtained for every tested voltage. 

\section{Conclusion}

In this paper, we have provided a practical study of two assembly-level software countermeasure against fault injection attacks. Even if those countermeasures are theoretically secure, it turns out that the level of security they add could be nullified if their implementation on a target platform is not performed in the right way. On this target platform with a variable-size instruction set, we need to make sure that no more than one instruction at a time is loaded at every clock cycle. This evaluation has also been performed on more complex codes from a FreeRTOS implementation.

The fault tolerance countermeasure has been very effective to protect an isolated subroutine call instruction. Thus, it seems such a sensitive instruction can be significantly reinforced against fault attacks. Yet, on a complex code, its results have been very mixed. Since this countermeasure has been formally proven resistant to instruction skips, its main limitation appears to be due to its considered fault model, which is probably too simplistic and does not provide a good enough coverage of the induced faults. A deeper understanding of the faults that can be produced in practice is necessary to define a more accurate fault model and then improve this countermeasure.

The fault detection countermeasure has been designed to protect a smaller set of instructions. The countermeasure has been very effective on the considered test cases. On a more complex code which only contains instructions for which a fault detection approach can be used, the countermeasure greatly increases the security level. However, its main drawback appears to be its limitation to a restricted set of instructions. Since the experimental results have shown a very good efficiency, this countermeasure needs to be extended to a larger set of instructions.

A possible extension of this work could evaluate the impact of such assembly-level countermeasures on the side-channel leakages of the circuit and study whether they can be combined with other software countermeasures against leakages.

\printbibliography[heading=bibintoc,title={References}]

\end{document}